\documentclass[sigconf,natbib=true,anonymous=false]{acmart}

\usepackage[hang,flushmargin]{footmisc}
\usepackage{xspace}
\usepackage{enumitem}
\usepackage{graphicx}
\usepackage{booktabs} 
\usepackage{xspace}
\usepackage{multirow}
\usepackage{caption}
\usepackage{subcaption}
\usepackage{amsmath}
\usepackage{booktabs, caption, makecell}
\usepackage{adjustbox}
\setlist{leftmargin=0mm}

\makeatletter
\newcommand\footnoteref[1]{\protected@xdef\@thefnmark{\ref{#1}}\@footnotemark}
\makeatother

\copyrightyear{2023}
\acmYear{2023}
\setcopyright{acmlicensed}
\acmConference[SIGIR '23]{Proceedings of the 46th International ACM SIGIR Conference on Research and Development in Information Retrieval}{July 23--27, 2023}{Taipei, Taiwan}
\acmBooktitle{Proceedings of the 46th International ACM SIGIR Conference on Research and Development in Information Retrieval (SIGIR '23), July 23--27, 2023, Taipei, Taiwan}
\acmPrice{15.00}
\acmDOI{10.1145/3539618.3591977}
\acmISBN{978-1-4503-9408-6/23/07}

\begin{document}

\title{SLIM: Sparsified Late Interaction for Multi-Vector Retrieval with Inverted Indexes}

\author{Minghan Li}
\affiliation{%
  \institution{University of Waterloo}
  \city{Waterloo}
  \country{Canada}
}
\email{m692li@uwaterloo.ca}
\author{Sheng-Chieh Lin}
\affiliation{%
  \institution{University of Waterloo}
  \city{Waterloo}
  \country{Canada}
}
\email{s269lin@uwaterloo.ca}
\author{Xueguang Ma}
\affiliation{%
  \institution{University of Waterloo}
  \city{Waterloo}
  \country{Canada}
}
\email{x93ma@uwaterloo.ca}
\author{Jimmy Lin}
\affiliation{%
  \institution{University of Waterloo}
  \city{Waterloo}
  \country{Canada}
}
\email{jimmylin@uwaterloo.ca}
\settopmatter{authorsperrow=4}

\renewcommand{\shortauthors}{Minghan Li, Sheng-Chieh Lin, Xueguang Ma, and Jimmy Lin}



\begin{abstract}
This paper introduces \textbf{S}parsified \textbf{L}ate \textbf{I}nteraction for \textbf{M}ulti-vector (SLIM) retrieval with inverted indexes. 
Multi-vector retrieval methods have demonstrated their effectiveness on various retrieval datasets, and among them, ColBERT is the most established method based on the late interaction of contextualized token embeddings of pre-trained language models.
However, efficient ColBERT implementations require complex engineering and cannot take advantage of off-the-shelf search libraries, impeding their practical use.
To address this issue, SLIM first maps each contextualized token vector to a sparse, high-dimensional lexical space before performing late interaction between these sparse token embeddings.
We then introduce an efficient two-stage retrieval architecture that includes inverted index retrieval followed by a score refinement module to approximate the sparsified late interaction, which is fully compatible with off-the-shelf lexical search libraries such as Lucene. 
SLIM achieves competitive accuracy on MS MARCO Passages and BEIR compared to ColBERT while being much smaller and faster on CPUs.
To our knowledge, we are the first to explore using sparse token representations for multi-vector retrieval.
Source code and data are integrated into the Pyserini IR toolkit.
\end{abstract}



\maketitle

\section{Introduction}
Pairwise token interaction~\cite{he-lin-2016-pairwise,lan-xu-2018-neural,Zhang2019ExplicitPW} has been widely used in information retrieval tasks~\cite{Manning2005IntroductionTI}.
Interaction-based methods enable deep coupling between queries and documents, which often outperform representation-based methods~\cite{reimers-gurevych-2019-sentence,cer2018universal} when strong text encoders are absent.
However, with the rise of pre-trained transformer models~\cite{devlin-etal-2019-bert,Liu2019RoBERTaAR}, representation-based methods such as DPR~\cite{karpukhin-etal-2020-dense} and SPLADE~\cite{Formal2021SPLADESL} gain more popularity as the pre-trained representations capture rich semantics of the input texts.
Moreover, these methods can leverage established search libraries such as FAISS~\cite{faiss} and Pyserini~\cite{lin2021pyserini} for efficient retrieval.
To combine the best of both worlds, models such as ColBERT~\cite{colbert} and COIL~\cite{gao-etal-2021-coil} that leverage late interaction are proposed, where the token interaction between queries and documents only happens at the last layer of contextualized embeddings.
Their effectiveness and robustness are demonstrated on various retrieval and question-answering benchmarks.
However, these models require token-level retrieval and aggregation, which results in large indexes and high retrieval latency.
Therefore, different optimization schemes are proposed to improve efficiency in both time and space~\cite{santhanam2022plaid,colberter}, making multi-vector retrieval difficult to fit into off-the-shelf search libraries.

In this paper, we propose an efficient and compact approach called \textbf{S}parsified \textbf{L}ate \textbf{I}nteraction for \textbf{M}ulti-vector retrieval with inverted indexes, or SLIM.
SLIM is as effective as the state-of-the-art multi-vector model ColBERT on MS MARCO Passages~\cite{nguyen2016ms} and BEIR~\cite{Thakur2021BEIRAH} without any distillation or hard negative mining while being more efficient.
More importantly, SLIM is fully compatible with inverted indexes in existing search toolkits such as Pyserini, with only a few extra lines of code in the pre-processing and post-processing steps.
In contrast, methods such as ColBERT and COIL require custom implementations and optimizations~\cite{santhanam2022plaid,colberter} in order to be practical to use.
This presents a disadvantage because some researchers and practitioners may prefer neural retrievers that are compatible with existing infrastructure based on inverted indexes so that the models can be easily deployed in production.


\begin{figure*}[t!]
\hspace{-0.1cm}
\begin{subfigure}[t]{.43\linewidth}
\centering
\includegraphics[width=1\textwidth]{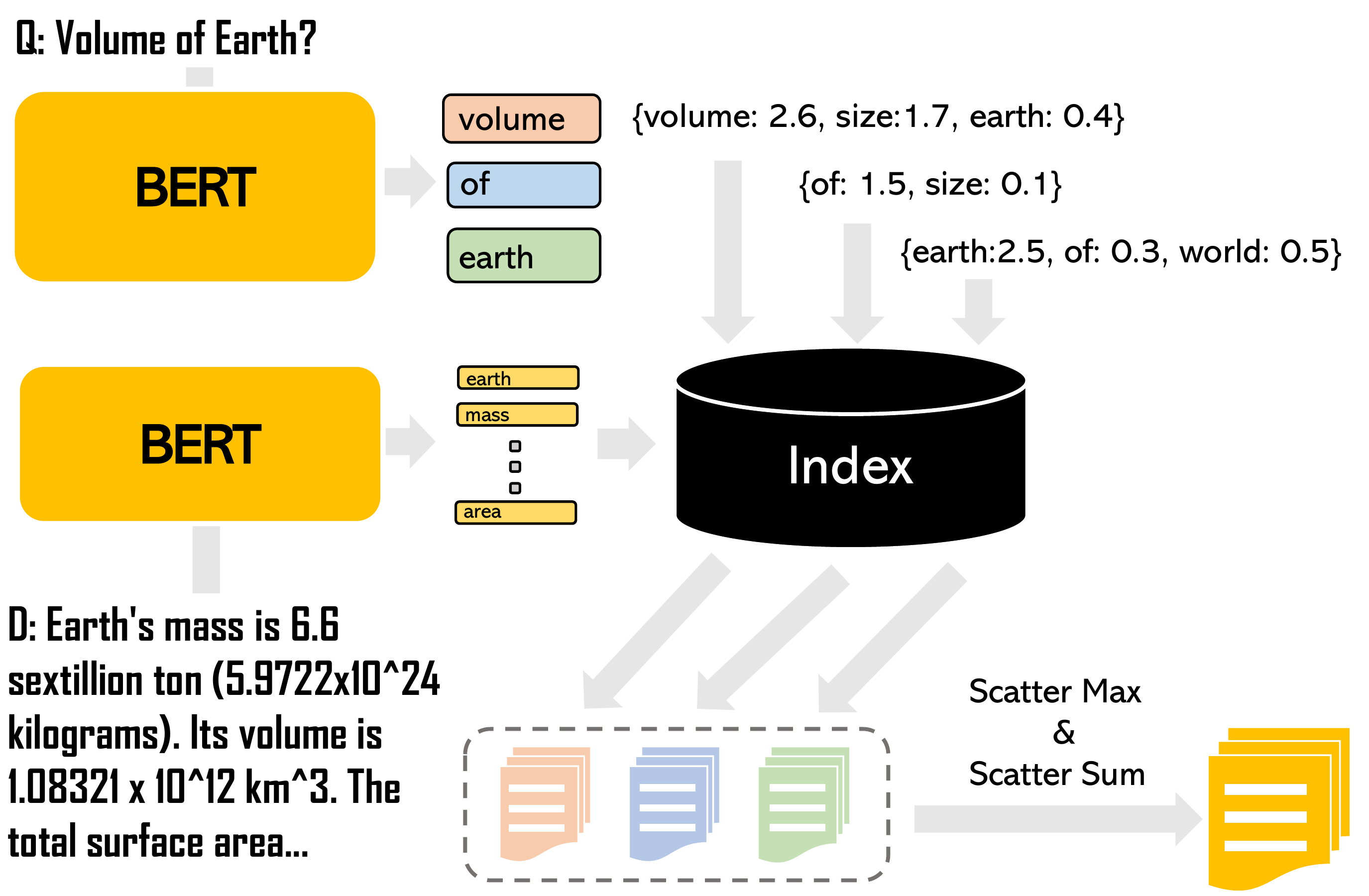}
\caption{Naive SLIM implementation.}
\label{fig:naive_slim}
\end{subfigure}%
\hspace{0.8cm} 
\begin{subfigure}[t]{.43\linewidth}
\centering
\includegraphics[width=1\textwidth]{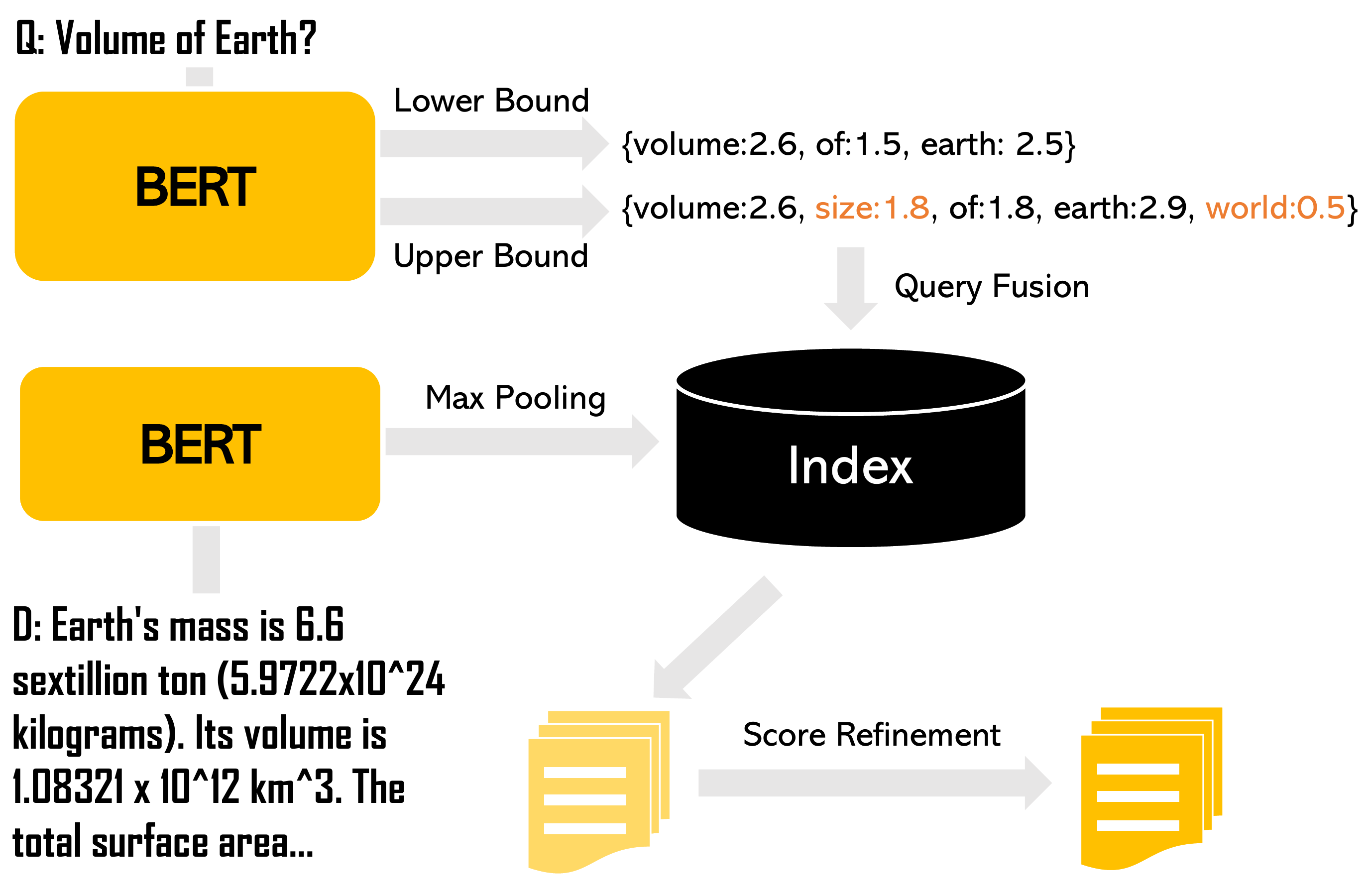}
\caption{Approximate SLIM implementation.}
\label{fig:app_slim}
\end{subfigure}\\
\caption{(a) The naive SLIM implementation takes the expansion of each token as sub-queries/documents. The token-level rankings are merged using scatter operations. (b) The approximate SLIM implementation first fuses the lower- and upper-bounds of scores for retrieval using Equations~\eqref{eq:factorization} and~\eqref{eq:slim_fusion}; the candidate list is then refined according to Equation~\eqref{eq:slim_sim}.}
\label{fig:slim}
\end{figure*}

In order to leverage standard inverted index search, SLIM projects each contextualized token embedding to a high-dimensional, sparse lexical space~\cite{Formal2021SPLADESL} before performing the late interaction operation.
Ideally, the trained sparse representations of documents can be compressed and indexed into an inverted file system for search.
Our method is the first to explore using sparse representations for multi-vector retrieval with inverted indexes to our knowledge.
However, unlike previous supervised sparse retrieval methods such as uniCOIL~\cite{unicoil} or SPLADE~\cite{Formal2021SPLADESL}, which compute sequence-level representations for each query/document, deploying token-level sparse vectors into inverted indexes is problematic.
As shown in Figure~\ref{fig:naive_slim}, the high-level idea is to convert each sparse token embedding into a sub-query/document and perform token-level inverted list retrieval before aggregation (e.g., scatter sum and max).
However, this is incredibly slow in practice as latency increases rapidly when the posting lists are too long.

Therefore, instead of using the naive approach, a more economical way is to approximate the sparsified late interaction with a two-stage system. 
We first calculate the upper and lower bounds of the sparsified late-interaction scores by swapping the order of the operators in late interaction (see Section~\ref{sec:method}).
In this way, the token-level sparse vectors are pooled before indexing to obtain sequence-level representations for the inverted index, yielding fast retrieval speed.
After retrieval, a lightweight score refinement step is applied, where we use the sparse token vectors stored using Scipy~\cite{2020SciPy-NMeth} to rerank the candidate lists.
As shown in Figure~\ref{fig:app_slim}, this two-stage design allows us to optimize the latency of the retrieval stage without worrying about accuracy loss as the score refinement will compensate for errors from the first stage.
Experiments on MS MARCO Passages show that SLIM is able to achieve a similar ranking accuracy compared to ColBERT-v2, while using $40\%$ less storage and achieving an $83\%$ decrease in latency.
To sum up, our contributions in this paper are three-fold:
\begin{itemize}[leftmargin=*]
    \item We are the first to use sparse representations for multi-vector retrieval with standard inverted indexes.
    \item We provide a two-stage implementation of SLIM by approximating late interaction followed by a score refinement step.
    \item SLIM is fully compatible with off-the-shelf search toolkits such as Pyserini.
\end{itemize}

\section{Methodology}\label{sec:method}
ColBERT~\cite{colbert} proposes late interaction between the tokens in a query $q=\{q_1, q_2,\cdots,q_N\}$ and a document $d=\{d_1, d_2,\cdots,d_M\}$:
\begin{align}\label{eq:colbert_sim}
    s(q,d) =  \sum_{i=1}^N \max_{j=1}^M v_{q_i}^T v_{d_j},
\end{align}
where $v_{q_i}$ and $v_{d_j}$ denote the last-layer contextualized token embeddings of BERT. 
This operation exhaustively compares each query token to all document tokens.
The latency and storage of ColBERT are bloated as many tokens do not contribute to either query or document semantics, and thus complex engineering optimizations are needed to make the model practical~\cite{santhanam-etal-2022-colbertv2,santhanam2022plaid}.

Unlike ColBERT, which only uses contextualized token embeddings for computing similarity, SPLADE~\cite{Formal2021SPLADESL,Formal2021SPLADEVS} further utilizes the pre-trained Mask Language Modeling (MLM) layer to project each $c$-dimensional token embedding to a high-dimensional, lexical space $V$.
Each dimension corresponds to a token and has a non-negative value due to the following activation:
\begin{align}\label{eq:splade_act}
    \phi_{d_j} = \log(1+\text{ReLU}(W^Tv_{d_j}+b)),
\end{align}
where $\phi_{d_j}\in \mathbb{R}^{|V|}$, $v_{d_j}$ is the token embedding of the $j$th token of document $d$; $W$ and $b$ are the weights and biases of the MLM layer.
To compute the final similarity score, SPLADE pools all the token embeddings into a sequence-level representation and uses the dot product between a query and a document as the similarity:
\begin{align}\label{eq:splade_sim}
    s(q,d) = (\max_{i=1}^{N} \phi_{q_i})^T (\max_{j=1}^{M} \phi_{d_j}).
\end{align}
Here, max pooling is an element-wise operation and is feasible for sparse representations as the dimensions of all the token vectors are aligned (lexical space) and non-negative.
These two properties play a vital role in the success of SPLADE and are also important for making SLIM efficient, as we shall see later.

\begin{table*}[t]
\centering
\begin{adjustbox}{max width=0.9\textwidth}
\begin{tabular}{l|cc|cc|cc|c|rr}
\toprule
\textbf{Models}& \multicolumn{2}{c|}{\textbf{MARCO Dev}}& \multicolumn{2}{c|}{\textbf{TREC DL19}}& \multicolumn{2}{c|}{\textbf{TREC DL20}} & \multicolumn{1}{c|}{\textbf{BEIR (13 tasks)}} & \multicolumn{2}{c}{\textbf{Space\&Time Efficiency}}\\
& \small{MRR@10} &\small{R@1k} &\small{nDCG@10} &\small{R@1k}&\small{nDCG@10} &\small{R@1k} &\small{nDCG@10} &\small{Disk (GB)} &\small{Latency (ms/query)}\\
\midrule
\multicolumn{9}{c}{\qquad \qquad \qquad \textit{Models trained with only BM25 hard negatives from MS MARCO Passages}}\\\midrule
BM25 & 0.188 &0.858 &0.506 &0.739 &0.488 &0.733 &0.440 &\textbf{\textbf{0.7}} & \textbf{\textbf{40}}\\
DPR &0.319 &0.941 &0.611 &0.742 &0.591 &0.796 &0.375 &26.0 &2015\\
SPLADE &0.340 &0.965 &0.683 &0.813 &0.671 &0.823 &0.453 &2.6 & 475\\
COIL &0.353 &0.967 &\textbf{0.704} &\textbf{0.835} &\textbf{0.688} &0.841 &\textbf{0.483} &78.5  &3258\\
ColBERT &\textbf{0.360} &\textbf{0.968} &0.694 &0.830 &0.676 &0.837  &0.453 &154.3 &-  \\
SLIM &0.358 &0.962 &0.701 &0.824 &0.640 &\textbf{0.854} &0.451 &18.2 &580\\
\midrule
\multicolumn{9}{c}{\qquad \qquad \qquad \textit{Models trained with further pre-training/hard-negative mining/distillation}}\\\midrule
coCondenser &0.382 &0.984 &0.674 &0.820  &0.684 &0.839 &0.420 &26.0 &2015\\
SPLADE-v2 & 0.368 &0.979 &0.729 &0.865 &0.718 &0.890 &0.499 &\textbf{4.1} &2710\\
ColBERT-v2 &0.397 &\textbf{\textbf{0.985}} &\textbf{\textbf{0.744}} &\textbf{\textbf{0.882}} &\textbf{\textbf{0.750}} &\textbf{\textbf{0.894}} &\textbf{\textbf{0.500}} &29.0 &3275\\
SLIM$^{++}$ &\textbf{\textbf{0.404}} &0.968 &0.714 &0.842 &0.702 &0.855 &0.490 & 17.3 &\textbf{550}\\
\bottomrule
\end{tabular}
\end{adjustbox}
\vspace{0.2cm}
\caption{In-domain and out-of-domain evaluation on MS MARCO Passages, TREC DL 2019/2020, and BEIR. ``-'' means not practical to evaluate on a single CPU. Latency is benchmarked on a single CPU and query encoding time is excluded.}
\vspace{-3mm}
\label{tbl:msmarco}
\end{table*}

Similar to ColBERT's late interaction, sparsified late interaction also takes advantages of the contextualized embeddings of BERT's last layer.
But different from ColBERT, we first apply the sparse activation in Equation~\eqref{eq:splade_act} before calculating the similarity in Equation~\eqref{eq:colbert_sim}:
\begin{align}\label{eq:slim_sim}
    s(q,d) =  \sum_{i=1}^N \max_{j=1}^{M} \phi_{q_i}^T \phi_{d_j},
\end{align}
where $\phi_{d_j}$ and $\phi_{q_i}$ are the representations in Equation~\eqref{eq:splade_act}.
However, as shown in Figure~\ref{fig:naive_slim}, to keep the token-level interaction, we must convert the sparse representation of each token in a query to a sub-query and retrieve the sub-documents from the index.
Moreover, the token-level rankings need to be merged to yield the final ranked list using scatter operations, which further increases latency.

Due to the impracticality of the naive implementation, a natural solution would be to approximate Equation~\eqref{eq:slim_sim} during retrieval. We first unfold the dot product in Equation~\eqref{eq:slim_sim}:
\begin{align}\label{eq:slim_sim_unfold}
s(q,d) = \sum_{i=1}^N \max_{j=1}^M \sum_{k=1}^{|V|} \phi_{q_i}^{(k)} \phi_{d_j}^{(k)},
\end{align}
where $\phi_{q_i}^{(k)}$ is the $k$th elements of $\phi_{q_i}$.
As each $\phi_{q_i}$ across different $q_i$ all shares the same lexical space and the values are non-negative, we can easily derive its upper-bound and lower-bound:
\begin{align}
    \sum_{i=1}^N\max_{j=1}^M \sum_{k=1}^{|V|} e_{q_i}^{(k)} \phi_{d_j}^{(k)} \leq s(q,d) \leq \sum_{i=1}^N \sum_{k=1}^{|V|} \max_{j=1}^M  \phi_{q_i}^{(k)} \phi_{d_j}^{(k)},  
\end{align}
\begin{align}
e_{q_i}^{(k)} =    \left\{\begin{matrix}
\phi_{q_i}^{(k)}, & \text{if $k=\underset{k}{\text{argmax }}\phi_{q_i}^{(k)}$} \\ 
0, & \text{otherwise}
\end{matrix}\right.
\end{align}
Then, the lower-bound score $s_l(q,d)$ and upper-bound score $s_h(q,d)$ can be further factorized as:
\begin{align}\label{eq:factorization}
     s_{l}(q,d)=\sum_{i=1}^N\max_{j=1}^M \sum_{k=1}^{|V|} e_{q_i}^{(k)} \phi_{d_j}^{(k)} &= (\sum_{i=1}^Ne_{q_i})^T(\max_{j=1}^M\phi_{d_j});\nonumber\\
    s_{h}(q,d)=\sum_{i=1}^N \sum_{k=1}^{|V|} \max_{j=1}^M  \phi_{q_i}^{(k)} \phi_{d_j}^{(k)} &= (\sum_{i=1}^N\phi_{q_i})^T(\max_{j=1}^M\phi_{d_j})
\end{align}
where both the sum and max operations will be element-wise if the targets are vectors.
We see that these two equations resemble the form of SPLADE in Equation~\eqref{eq:splade_sim}, where queries and documents are encoded into sequence-level representations independently.
In this way, $\max_{j=1}^M\phi_{d_j}$ in Equation~\eqref{eq:factorization} can be pre-computed and indexed offline.
To approximate $s(q,d)$ in Equation~\eqref{eq:slim_sim_unfold}, we use the linear interpolation between the lower- and the upper-bound:
\begin{align}\label{eq:slim_fusion}
    s_{a}(q,d) &= \beta\cdot s_l(q,d) + (1-\beta)\cdot s_h(q,d)\nonumber\\
    &=(\sum_{i=1}^N(\beta\cdot e_{q_i} + (1-\beta)\cdot \phi_{q_i}))^T(\max_{j=1}^M\phi_{d_j})
\end{align}
where $s_a(q,d)$ is the approximate score of SLIM and $\beta\in [0,1]$ is the interpolation coefficient.
As shown in Equation~\eqref{eq:slim_fusion}, we can first fuse the query representation before retrieval and fine-tune the coefficient $\beta$ on a validation set.
It is worth mentioning that the approximation is applied after the SLIM model is trained using Equation~\eqref{eq:slim_sim}.
During indexing, we use Pyserini~\cite{lin2021pyserini} to index the sequence level-representation $\max_{j=1}^M\phi_{d_j}$ and use Scipy~\cite{2020SciPy-NMeth} to store the sparse token vectors.
During retrieval, we use Equation~\eqref{eq:slim_fusion} to retrieve the top-$k$ candidates with the \texttt{LuceneImpactSearcher} in Pyserini.
For score refinement, we extract the stored sparse token vectors for the top-$k$ documents and use Equation~\eqref{eq:slim_sim} to refine the candidate list.
This two-stage breakdown yields a good effectiveness-efficiency trade-off, as we can aggressively reduce the first-stage retrieval latency without too much accuracy loss due to the score refinement.

To further reduce the memory footprint and improve latency for the first-stage retrieval, we apply two post-hoc pruning strategies on the inverted index to remove:
(1) tokens that have term importance below a certain threshold;
    (2) postings that exceed a certain length (i.e., the inverse document frequency is below a threshold).
Since all the term importance values are non-negative, tokens with smaller weights contribute less to the final similarity.
Moreover, tokens with long postings mean that they frequently occur in different documents, which yield low inverse document frequencies (IDFs) and therefore can be safely discarded.

\begin{figure*}[t!]
\begin{subfigure}[t]{.3\linewidth}
\centering
\includegraphics[width=1\textwidth]{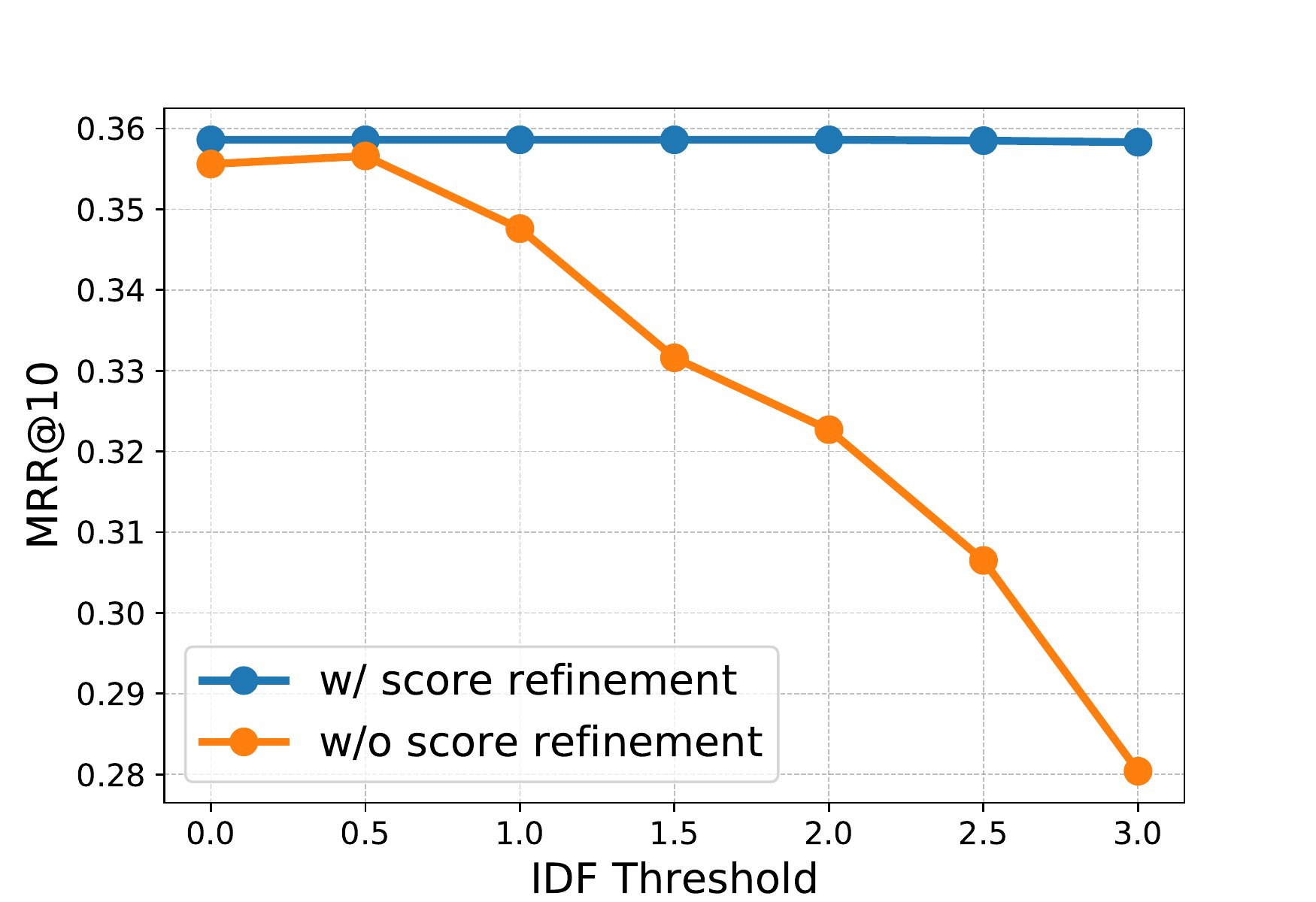}
\caption{MRR@10 vs IDF.}
\end{subfigure}%
\begin{subfigure}[t]{.3\linewidth}
\centering
\includegraphics[width=1\textwidth]{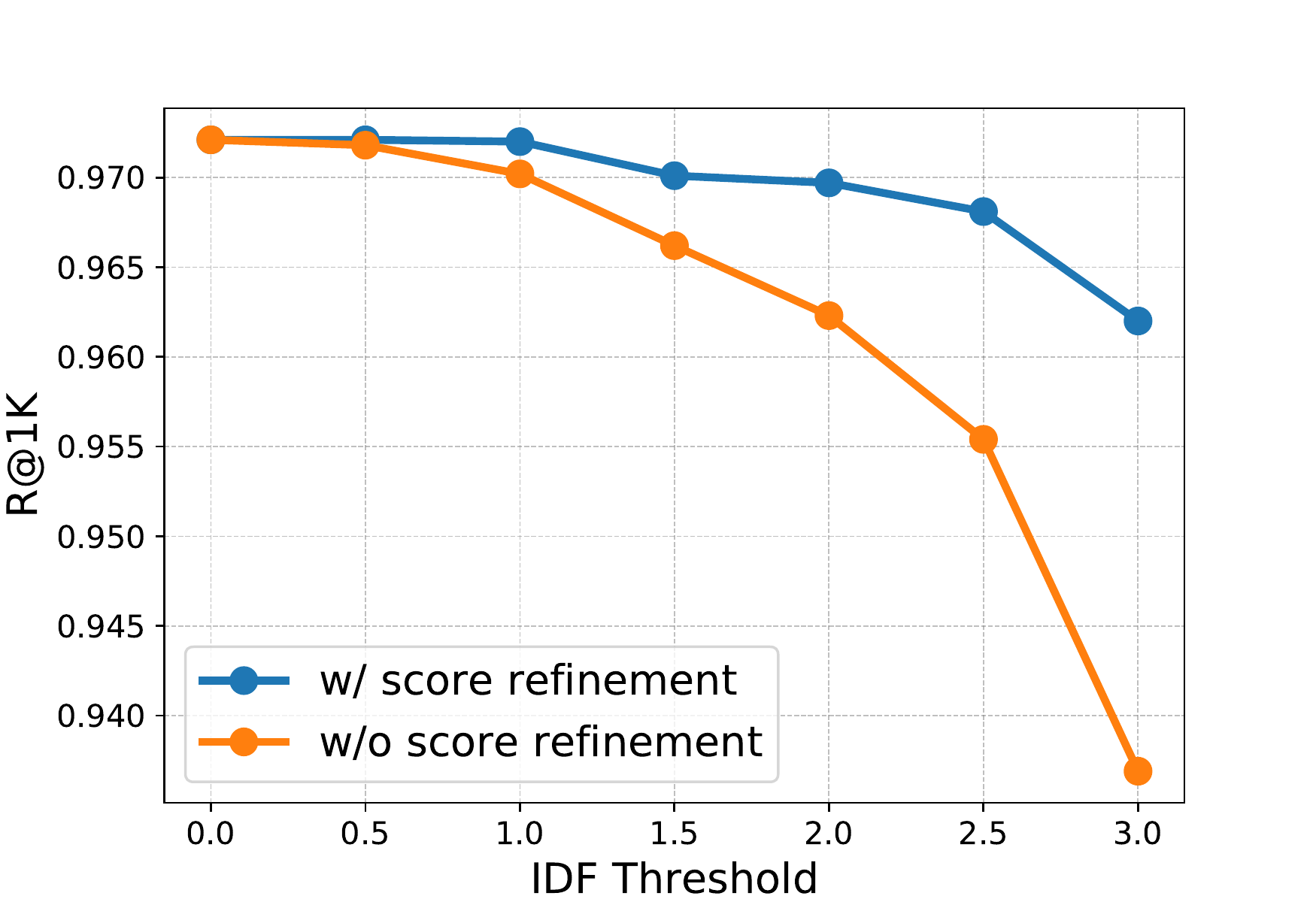}
\caption{Recall@1000 vs IDF.}
\end{subfigure}
\begin{subfigure}[t]{.3\linewidth}
\centering
\includegraphics[width=1\textwidth]{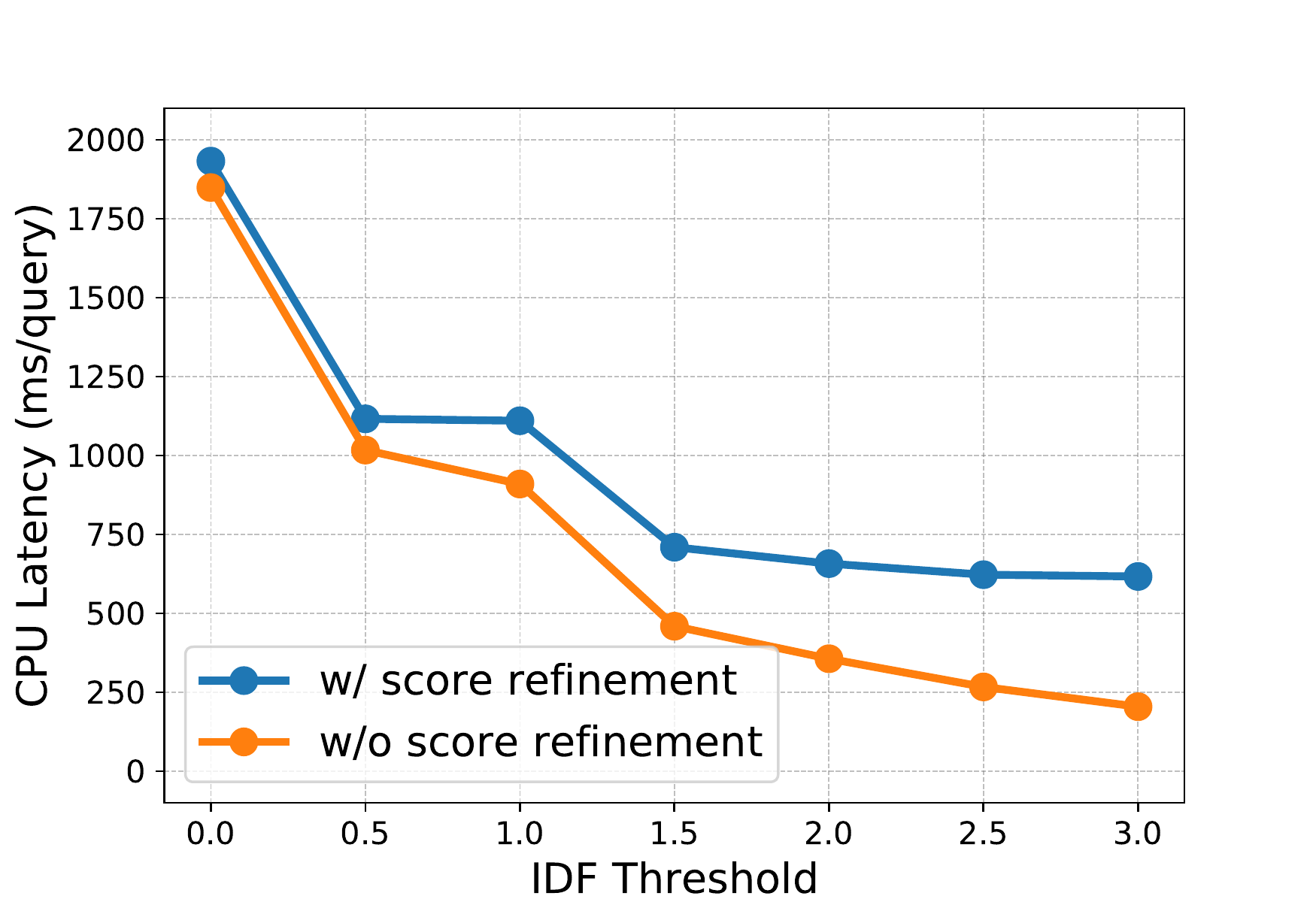}
\caption{CPU Latency vs IDF.}
\end{subfigure}\\
\caption{Effectiveness-Efficiency trade-offs of SLIM (w/ and w/o score refinement) on MS MARCO Passages.}
\label{fig:tradeoff}
\end{figure*}

\section{Experiments}\label{sec:exp}


We evaluate our model and baselines on MS MARCO Passages~\cite{nguyen2016ms} and its shared tasks, TREC DL 2019/2020 passage ranking tasks~\cite{craswell2020overview}.
We train the baseline models on MS MARCO Passages and report results on its dev-small set and TREC DL 2019/2020 test queries following the same setup in CITADEL~\cite{citadel}.
We further evaluate the models on the BEIR benchmark~\cite{Thakur2021BEIRAH}, which consists of a diverse set of 18 retrieval tasks across 9 domains.  
Following previous work~\cite{santhanam-etal-2022-colbertv2,Formal2021SPLADEVS}, we only evaluate 13 datasets due to license restrictions.\footnote{CQADupstack, Signal-1M(RT), BioASQ, Robust04, TREC-NEWs are excluded.}
The evaluation metrics are MRR@$10$, nDCG@$10$, and Recall@$1000$ (i.e., R@1K).
Latency is benchmarked on a single Intel(R) Xeon(R) Platinum 8275CL CPU @ 3.00GHz and both the batch size and the number of threads are set to 1.

We follow the training procedure of CITADEL~\cite{citadel} to train SLIM and apply $\ell_1$ regularization on the sparse token representations for sparsity control.
SLIM is trained only with BM25 hard negatives in MS MARCO Passages, while SLIM$^{++}$ is trained with cross-encoder distillation and hard negative mining.
For indexing and pruning, the default token weight pruning threshold is $0.5$ and the IDF threshold is $3$.
We use Pyserini to index and retrieve the documents and use Scipy to store the sparse token vectors (CSR matrix).
For retrieval, the fusion weight $\beta$ in Equation~\eqref{eq:slim_fusion} is set to $0.01$.
We first retrieve the top-$4000$ candidates using Pyserini's \texttt{LuceneImpactSearcher} and then use the original sparse token vectors stored by Scipy to refine the ranked list and output the top-$1000$ documents.

Table~\ref{tbl:msmarco} shows the in-domain evaluation results on MS MARCO Passages and TREC DL 2019/2020.
SLIM and SLIM$^{++}$ manage to achieve accuracy comparable to ColBERT and ColBERT-v2 on MS MARCO Passages.
However, the results seem unstable on TREC DL, which is expected as TREC DL has far fewer queries.
For out-of-domain evaluation, the results are a little bit worse than the current state of the art but still close to ColBERT.
For latency and storage, SLIM has a much smaller disk requirement and lower latency compared to ColBERT.
For example, SLIM$^{++}$ achieves effectiveness comparable to ColBERT-v2 with an $83\%$ decrease in latency on a single CPU and using $40\%$ less storage.


Next, we show that a two-stage implementation of SLIM provides a good trade-off between effectiveness and efficiency.
Figure~\ref{fig:tradeoff} plots CPU latency vs.\ MRR@$10$/Recall@1000 on MS MARCO Passages using IDF thresholding for SLIM (w/o hard-negative mining and distillation).
We set the minimum IDF threshold to $\{$0, 0.5, 1, 1.5, 2, 2.5, 3$\}$ and first-stage retrieval candidates top-$k$ to $\{$1000, 1500, 2000, 2500, 3000, 3500, 4000$\}$ before score refinement (if any).
Without the score refinement, the MRR@10 and Recall@1000 scores are high when the IDF threshold is small, but effectiveness drops rapidly when we increase the IDF threshold to trade accuracy for latency.
In contrast, with the score refinement step, we see that with about $0.003$ loss in MRR@$10$ and $0.01$ loss in Recall@$1000$, latency improves drastically from about $1800$ ms/query to $500$ ms/query.
The reason is that the score refinement step, which takes a small amount of time compared to retrieval, compensates for the errors from the aggressive pruning at the first stage, which shows that our novel lower- and upper-bound fusion provides a good approximation of SLIM when the IDF threshold is low, and that the score refinement step is important as it allows more aggressive pruning for the first-stage retrieval.

\section{Related Work}\label{sec:related}

Dense retrieval~\cite{karpukhin-etal-2020-dense} gains much popularity as it is supported by multiple approximate nearest neighbor search libraries~\cite{faiss,scann}.
To improve effectiveness, hard negative mining~\cite{ance2021xiong,star} and knowledge distillation~\cite{lin-etal-2021-batch, tasb} are often deployed. 
Recently, further pre-training for retrieval~\cite{gao-etal-2021-simcse, lu-etal-2021-less, gao-callan-2021-condenser, contriever, gao-callan-2022-unsupervised} is proposed to improve the fine-tuned effectiveness of downstream tasks. 

Sparse retrieval systems such as BM25~\cite{robertson2009bm25} and tf--idf~\cite{Salton_Buckley_1988} encode documents into bags of words. 
Recently, pre-trained language models are used to learn contextualized term importance~\cite{sparterm, deepimpact,Formal2021SPLADESL,unicoil,dragon}.
These models leverage current search toolkits such as Pyserini~\cite{lin2021pyserini} to perform sparse retrieval, or contribute to hybrid approaches with dense retrieval~\cite{colberter, unifier, dhr, aggretriever}.

Besides ColBERT~\cite{colbert}, COIL~\cite{gao-etal-2021-coil} accelerates retrieval by combining exact match and inverted index search.
CITADEL~\cite{citadel} further introduces lexical routing to avoid the lexical mismatch issue in COIL.
ME-BERT~\cite{luan-etal-2021-sparse} and MVR~\cite{zhang-etal-2022-multi} propose to use a portion of token embeddings for late interaction. 
ALIGNER~\cite{Qian2022MultiVectorRA} frames multi-vector retrieval as an alignment problem and uses entropy-regularized linear programming to solve it.

\section{Conclusion}
In this paper, we propose an efficient yet effective implementation for multi-vector retrieval using sparsified late interaction, which yields fast retrieval speed, small index size, and full compatibility with off-the-shelf search toolkits such as Pyserini.
The key ingredients of SLIM include sparse lexical projection using the MLM layer, computing the lower- and upper-bounds of the late interaction, as well as token weight and postings pruning.
Experiments on both in-domain and out-of-domain information retrieval datasets show that SLIM achieves comparable accuracy to ColBERT-v2 while being more efficient in terms of both space and time.

\section*{Acknowledgements}
This research was supported in part by the Natural Sciences and Engineering Research Council (NSERC) of Canada; computational resources were provided by Compute Canada.

\bibliographystyle{ACM-Reference-Format}
\bibliography{anthology,custom}

\appendix
\end{document}